\newcommand{\be}{\begin{equation}}
\newcommand{\ee}{\end{equation}}
\newcommand{\bea}{\begin{eqnarray}}
\newcommand{\eea}{\end{eqnarray}}
\newcommand{\bm}{\bibitem}

\newcommand{\bet}{\beta}
\newcommand{\gm}{\gamma}
\newcommand{\gz}{\gamma_0}
\newcommand{\gf}{\gamma_5}
\newcommand{\Gm}{\Gamma}
\newcommand{\ep}{\epsilon}

\newcommand{\de}{\delta}
\newcommand{\om}{\omega}
\newcommand{\omk}{\omega_k}
\newcommand{\omp}{\omega_p}

\newcommand{\tpp}{\theta (p_0)}
\newcommand{\tmp}{\theta (-p_0)}
\newcommand{\lm}{\lambda}

\newcommand{\sg}{\sigma}

\newcommand{\dm}{\bigtriangleup m}

\newcommand{\bn}{\overline{n}}
\newcommand{\obf}{\overline{f}}
\newcommand{\bta}{\overline{\eta}}
\newcommand{\bM}{\overline{M}}
\newcommand{\us}{u \!\!\! /}

\newcommand{\ps}{p \!\!\! /}

\newcommand{\Qs}{Q \!\!\!\! /}

\newcommand{\vq}{\vec{q}} 

\newcommand{\la}{\langle}
\newcommand{\ra}{\rangle}
\newcommand{\rw}{\rightarrow}

\newcommand{\f}{F_\pi}
\newcommand{\F}{F^2_{\pi}}

\documentclass[prd,aps,floats,nofootinbib,tightenlines,showpacs]{revtex4}
\usepackage{graphicx}
\begin{document}

\title{Nucleon propagation through nuclear matter in chiral effective field
theory}

\author{S. Mallik} 
\email{mallik@theory.saha.ernet.in}
\affiliation{Saha Institute of Nuclear Physics,
1/AF, Bidhannagar, Kolkata-700 064, India} 

\author{Hiranmaya Mishra} 
\email{hm@prl.res.in}
\affiliation{Theoretical Physics and Complex Systems Divison, 
Physical Research Laboratory, Ahmedabad 380 009, 
India}
\date{\today} 

\begin{abstract} 

We treat the propagation of nucleon in nuclear matter by evaluating the
ensemble average of the two-point function of nucleon currents in the 
framework of the chiral effective field theory. We first derive the effective 
parameters of nucleon to one loop. The resulting formula for the effective mass 
was known previously and gives an absurd value at normal nuclear density. We 
then modify it following Weinberg's method for the two-nucleon system in the 
effective theory. Our results for the effective mass and the width of nucleon 
are compared with those in the literature.

\end{abstract}


\maketitle
\section{introduction}

Chiral perturbation theory is the effective theory of strong interactions,
based only on the approximate chiral symmetry of the QCD Lagrangian and its
assumed spontaneous breaking. While this theory is eminently successful in
dealing with pionic processes \cite{Weinberg1,Gasser}, and also to a
great extent, with those of pions and a single nucleon \cite{Becher}, its
direct application to processes involving two or more nucleons is
problematic \cite{Epelbaum}. As first pointed out by Weinberg \cite{Weinberg2}, 
the contribution of graphs for such processes grow with increasing number of 
loops, in gross violation to the power counting rule. This failure of power 
counting is, of course, only to be expected, as precisely the sum 
over such (infinite number of) loop graphs would give rise to the bound states, 
such as the deuteron and the virtual bound state in the two nucleon system.

A better analysis of graphs can be made in the old fashioned perturbation 
theory \cite{Weinberg2}. Here it is the loop diagrams containing one or more 
pure-nucleon intermediate states that lead to divergence of the perturbation 
expansion, while those with intermediate states containing at least one pion 
obey the power counting rule. Accordingly Weinberg \cite{Weinberg2} proposes 
to construct first the effective potential from those connected graphs of the 
$T$-matrix, that do not contain any pure-nucleon intermediate state. Then the 
graphs with pure-nucleon intermediate states are summed over by the
Lippmann-Schwinger intergal equation for the $T$- matrix with this effective
potential.

In this work we apply this procedure to find the effective mass of nucleon 
in nuclear matter. The one-loop formula is long known \cite{Montano}, giving 
the mass-shift as the nucleon number density times a constant, depending on 
the parameters of the effective Lagrangian to leading order. When expressed in 
terms of $N\!N$ scattering lengths, this constant and hence the mass-shift
becomes too large to be acceptable.

What goes wrong with the one-loop result is clearly the approximation of the
appropriate $N\!N$ scattering amplitude, that is present in the formula, by 
merely the s-wave scattering lengths. To improve this amplitude within the
effective theory, the Weinberg analysis suggests that we regard the constant 
$N\!N$ amplitude as the effective potential to leading order and solve
the Lippmann-Schwinger equation for the complete amplitude. Replacing the constant 
amplitude in the one-loop formula by this solution, we get a greatly reduced 
value for the effective nucleon mass. We also find the effect of including
phenomenologically the effective ranges in the formula \cite{Kaplan}.

We first derive in some detail the old result for the nucleon effective
mass, using the real time formulation of field theory in matter
\cite{Landsmann}. We begin with the ensemble average of the two-point function 
of nucleon currents \cite{Ioffe}. It is evaluated to one-loop with vertices 
from the effective Lagrangian \cite{Weinberg2}, using the method of external
fields\cite{Gasser}. The nucleon pole term gives immediately the the
effective mass to leading order in the usual power counting. (We also get the
residue at the pole, but it contains an unknown coupling constant.) The constant 
scattering amplitude contained in it is then improved upon as described above.

In Sec.II we obtain the pieces of the effective Lagrangian, that will be
needed in evaluating the one-loop graphs. We use the Dirac, rather than the
Pauli, spinor for the nucleon. The advantage of this relativistic treatment 
is the clear separation of the vacuum and the density dependent parts of the 
nucleon propagator, the latter part containing the on-shell delta-function.  
In Sec.III we evaluate the graphs to obtain the leading terms for the effective 
parameters of nucleon in nuclear matter. We modify the mass shift formula 
in Sec.IV. Finally Sec.V includes some comments on our work. In Appendix A we 
describe a dispersion integral method to obtain the density dependent 
contributions from the one-loop graphs. The short Appendix B gives the result 
of partial wave expansion of the spin averaged $N\!N$ scattering amplitude in 
the forward direction.
   
\section{chiral lagrangian}
\setcounter{equation}{0}
\renewcommand{\theequation}{2.\arabic{equation}} 
 
We study the nucleon propagation in nuclear matter by analysing the two-point
function of the nucleon current $\eta (x)$ \cite{Ioffe}, built out of three quark
fields so as to have the quantum numbers of the nucleon. In this work we 
do not need to spell out the form of this current; it suffices only to note
its tranformation under the chiral symmetry group $SU(2)_R\times SU(2)_L$,
\be
\eta_R\rw g_R\eta_R,~~~~~\eta_L\rw g_L\eta_L, ~~~~~~~~~~~g_R\in SU(2)_R,~~~~~
g_L\in SU(2)_L,
\ee
where the subscripts $R$ and $L$ on $\eta$ denote its right- and the left-handed 
components, $\eta_{R,L}=\frac{1}{2}(1\pm \gf)\eta $.

To get the low energy structure of vertices with the nucleon current in the
effective theory, we couple it to an external (spinor) field $f(x)$, thereby 
extending the original $QCD$ Lagrangian ${\cal L}_{QCD}^{(0)}$ to
\be
{\cal L}_{QCD} = {\cal L}_{QCD}^{(0)} + \obf\eta + \bta f.
\ee
Writing in terms of $R$ and $L$ components,
\[\obf\eta =  \obf_R\eta_L +  \obf_L\eta_R , \]
we see that such a term is chirally invariant if the external fields
$f(x)_{R,L}$ transform oppositely to $\eta(x)_{R,L}$,
\be
f_R\rw g_L f_R,~~~~~f_L\rw g_R f_L\,. 
\ee
The transformation rules for the pion and the nucleon fields are known
\cite{Gasser,Ecker}. One can define three different quantities involving the 
pion triplet, namely $U,~~u~(u^2=U)$ and 
$u_\mu =iu^{\dag} (\partial_\mu u) u^{\dag}$ transforming as,
\begin{eqnarray}
U & \rw &  g_R U g_L^{\dag}, \nonumber \\
u &\rw & g_R u h^{\dag} = h u g_L^{\dag}, \nonumber \\
u_\mu & \rw & h u_\mu h^{\dag}\,,
\end{eqnarray} 
where $h$ is an element of the unbroken subgroup, $h \in SU(2)_V$. The
nucleon field $\psi (x)$ transforms according to the isospin $\frac{1}{2}$
representation of $SU(2)_V$,
\[\psi \rw h\psi.\]

We can now construct the effective Lagrangian for the nucleon-nucleon system 
including pions in presence of the external field $f$. Its terms may be put 
into three groups,
\be
{\cal L}_{e\!f\!f}  =  {\cal L}_{\pi \psi} + {\cal L}_{\psi^4} + {\cal
L}_{f}\,, 
\ee
which we now write down one by one. $ {\cal L}_{\pi \psi}$ is given by the 
familiar terms,
\be
{\cal L}_{\pi \psi} = \frac{\F}{4} \partial_\mu U \partial^\mu U^{\dag} +
\bar{\psi}(i {\partial \!\!\! /} -m)\psi +\frac{1}{2}g_A\bar{\psi}\us\gf\psi\,.
\ee
Here $g_A$ is the constant ($g_A=1.26$) appearing in the neutron beta-decay.
We choose the  explicit representation $U=\exp(i\phi^a\tau^a/\f)$,
where $\phi^a$ are the hermitian pion fields ($a=1,2,3$), $\tau^a$, the Pauli
matrices and $F_\pi$, the pion decay constant ($F_\pi=93$ MeV). In the 
following we do not need vertices with pion fields only.

The piece ${\cal L}_{\psi^4}$ giving the leading quartic
interaction of nucleons has been written by Weinberg as \cite{Weinberg2},
\[{\cal L}_{\psi^4}=-\frac{C_S}{2} (N^{\dag} N)^2 
-\frac{C_T}{2}(N^{\dag}\vec{\sg} N)^2 +\cdots, \]
where $C_S$ and $C_T$ are constants and $N(x)$ is the Pauli spinor (and
isospinor) for the nucleon and $'\cdots'$ refers to terms with two and more
derivatives. As we intend to work with the Dirac spinor 
$\psi (x)$, we rewrite it as  
\be
{\cal L}_{\psi^4}=-\frac{C_S}{8}\{\bar\psi (1+\gm_0)\psi\}^2
-\frac{C_T}{8}\{\bar\psi (1+\gz)\vec{\gm}\gf\psi\}^2 +\cdots\,.
\ee

In the piece ${\cal L}_f$ dependent on the external field, we need two couplings,
namely, that are linear and cubic in $\psi$,
\[{\cal L}_f ={\cal L}_{f\psi} +{\cal L}_{f\psi^3}.\]
Using the above transformation rules, the piece linear in $\psi$ may be written 
as
\begin{eqnarray}
{\cal L}_{f\psi} &=&
\lm \overline{(uf_R)}\psi +\lm'\overline{(u^{\dag} f_L)}\psi +h.c. \nonumber \\
&=& \frac{\lm}{2}\obf (1-\gf) u^{\dag} \psi +\frac{\lm'}{2}\obf
(1+\gf) u\psi +h.c.,
\end{eqnarray}
where invariance under parity requires $\lm=\lm'$. Following the construction 
of ${\cal L}_{\psi^4}$, we write the other piece as
\be
{\cal L}_{f\psi^3} =\frac{A_S}{4}\bar{\psi} (1+\gz)\psi \obf (1+\gm_0)
\psi +\frac{A_T}{4}\bar{\psi} (1+\gz)\vec{\gm}\gf\psi \obf (1+\gz) 
\vec{\gm}\gf \psi + h.c. 
\ee
Two other possible non-derivative terms involving $\vec{\tau}$ do not appear here for
the same reason as they did not in Eq.(2.6), namely that the
anticommutativity of the nucleon field allows each of these terms to be
written as linear combinations of the two terms written above. Clearly here
also there are terms with two or more derivatives on the nucleon field.

\section{one-loop formula}
\setcounter{equation}{0}
\renewcommand{\theequation}{3.\arabic{equation}} 

We start with the ensemble average of the two-point function of nucleon
currents in symmetric nuclear matter,
\be
\Pi (q)=i\int d^4x e^{iqx} \la T\eta (x) \bar{\eta} (0)\ra ,
\ee
where for any operator $O$, $\la O\ra= Tr\left[ \rho O \right]/Tr \rho~,~ 
\rho = e^{-\beta(H-\mu N)}$, $H$ being the Hamiltonian of 
the system, $\beta$ the inverse of temperature and $N$ the nucleon number 
operator with chemical potential $\mu$. In general, we deal here with
functions of two variables, namely $q_0\equiv E$ and $|\vq|$.

In the real time version of quantum field theory in a medium,  
the  two point function assumes the form of a $2\times 2$ matrix. But the dynamics 
is given essentially by a single analytic function, that is determined by 
the $11$-component itself. Denoting the $11$-component of the matrix
amplitude by $F_{11}(E,\vq)$, the corresponding analytic function $F(E,\vq)$ has the 
spectral representation in $E$ at fixed $\vq$ \cite{Landsmann},
\be
F(E,\vq)=\frac{1}{\pi}\int \frac{\coth(\bet (E'-\mu)/2) Im F_{11}(E',\vq)}
{E'-E-iE'\ep} dE'.
\ee
where the integral runs over cuts of $F_{11}(E,\vq)$.

The simplest example of a two-point function in nuclear medium is, of course, the
free nucleon propagator. Its $11$-component is given by \cite{Kobes},
\footnote{Going over to the non-relativistic limit and omitting anti-particle
contribution, it reduces to the more familiar form \cite{Fetter},
\[-S_{11}(p)=\frac{\theta (p-p_F)}{p_0-\om_p+i\ep} +\frac{\theta
(p_F-p)}{p_0-\om_p-i\ep}\,, \]
where the vacuum and the medium contributions are mixed up.}
\be
\frac{1}{i} S(p)_{11} \equiv \int d^4x e^{ipx}\la T\psi(x)\bar\psi(0)\ra_{11}
=(\ps+m)\left[\frac{i}{p^2-m^2+i\ep}-\left\{n^-(\omp)\tpp+
n^+(\omp)\tmp\right\}2\pi\delta(p^2-m^2)\right],
\ee
where $n^{\mp}$ are the distribution functions for nucleons and
antinucleons, $ n^\mp(\omp)=\left\{e^{\beta(\omp \mp\mu)} +1\right\}^{-1},
~\omp=\sqrt{m^2+p^2}$. The corresponding analytic function $S(p)$, as 
given by Eq. (3.2) is
\[S(p)=- (\ps+m)\frac{1}{p^2-m^2+i\ep}, \]
which is independent of $n^{\mp}$ and identical to the free propagator in vacuum. 

In the following we shall work in the limit of zero temperature, when the
distribution functions become, $n^- \rw \theta (\mu -\omp),~~n^+\rw 0$.
Then the nucleon number density, $\bn$ is given by
\[ \bn =4\int \frac{d^3p}{(2\pi)^3} \theta (\mu-\omp) =
\frac{2p_F^3}{3\pi^2},\]
where $p_F$ is the Fermi momentum related to the chemical potential $\mu$ by
$\mu^2=m^2+p_F^2$. We want to calculate the density dependent part of $\Pi(q)$ 
in the low energy region to first order in $\bn$. To this end we draw all the 
Feynman graphs with one loop containing a nucleon line. In addition to the 
single nucleon line forming a loop, we include also the loop containing an 
additional pion line to account for singularities with the lowest threshold. 
They are depicted in Fig. 1 along with the free propagator graph (a).

\begin{figure}
\begin{center}
\includegraphics[width=12cm,height=8cm]{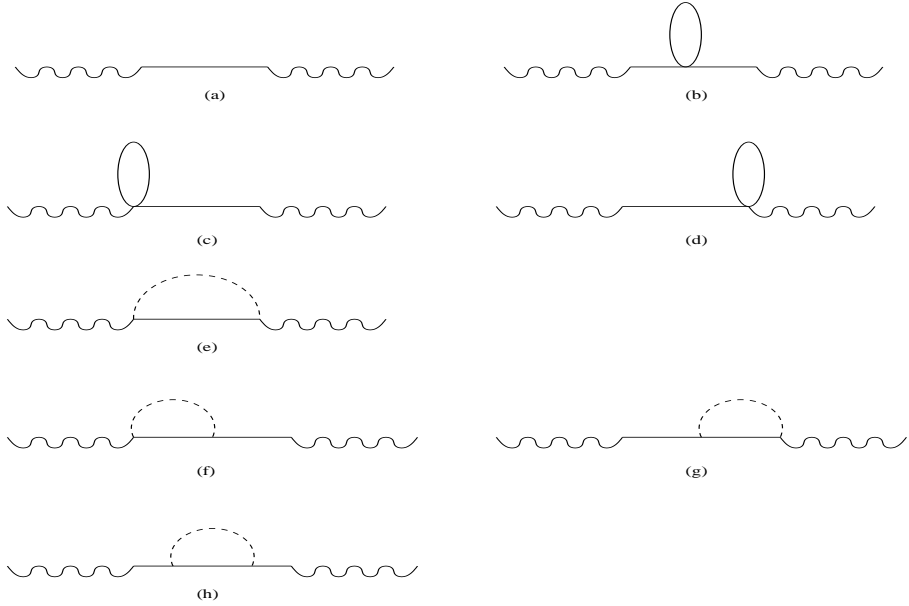}
\end{center}
\caption{Free propagator and one loop Feynman graphs for the two point 
function in the low energy region. The wavy line represents the nucleon 
current, while the continuous and the dashed internal lines are for nucleon
and pion propagation.}
\end{figure}

Before we work out the graphs, it is possible to simplify the piece 
${\cal L}_{f\psi^3}$ in the Lagrangian, needed for vertices in graphs (c)
and (d) of Fig.1. As two of the nucleon fields are contracted at the vertex
itself, we can do so already in ${\cal L}_{f\psi^3}$ and retain only the density 
dependent part in the contraction.Thus referring to Eq.(2.2), we get from
Eqs. (2.8 - 9) the complete expression for $\eta$ to leading order in the 
effective theory as, 
\be
\eta=\lm\left\{ 1 +\frac{i\pi\cdot\tau}{2\f}\gf +\zeta\,\bn (1+\gm_0)
\right\} \psi, ~~~~ \zeta=\frac{3}{8\lm} (A_S-A_T).
\ee
We shall work at $\vq=0$, when the free propagation graph 
(a) takes the form 
\be
\Pi(E)_{(a)}= \lm^2 S(E) = -\frac{\lm^2}{E - m + i\ep}\,\frac{1}{2} (1+\gz),
\ee
in the vicinity of the nucleon pole. We now calculate the corrections to it
produced by the rest of the graphs.

To evaluate the self-energy diagram (b), we rewrite the four-nucleon 
interaction term given by Eq. (2.6) conveniently as
\be
{\cal L}_{\psi^4}=-\frac{1}{8}\sum_{i=S,T}C_i\Gm^i_{AB}\Gm^i_{CD}
(\bar\psi_A\psi_B)(\bar\psi_C\psi_D),
\ee
where $\Gm^S=(1+\gm_0)$ and $\Gm^T=(1+\gz)\vec\gm\gf$.
Then the two point function for this diagram is given by
\be
\Pi(E)_{(b)}=-\lm^2 S(E) \sg S(E).
\ee
Here the self-energy $\sg$ is a constant given by
\be
\sg=-\frac{i}{4}\sum_{i=S,T}C_i\int\frac{d^4p}{(2\pi)^4} 
\{-2tr(S(p)_{11}\Gm^i)\Gm^i+\Gm^iS(p)_{11}\Gm^i\},
\ee
where the $tr$(ace) is over $\gm$-matrices. On inserting the density 
dependent part of $S_{11}$ from Eq. (3.3), $\sg$ can be immediately
evaluated to give
\be
\Pi(E)_{(b)}=-\frac{3\lm^2}{4}(C_S-C_T)\bn\frac{1}{(E-m)^2}\frac{1}{2}(1+\gz).
\ee
Note the absence of a simple pole in this contribution. From the constant
vertex graphs (c) and (d), we get
\be
\Pi (E)_{(c)+(d)}= -4\lm^2\zeta\, \bn\frac{1}{E-m}\frac{1}{2}(1+\gz).
\ee

The graph (e) has the two-particle ($\pi$ and $N$) intermediate state. 
It is evaluated in detail in Appendix A. As expected, it does not give 
rise to a pole at $E=m$. Following similar steps, we get the 
contribution of the remaining graphs as
\bea
\Pi (E)_{(f)+(g)} &=& \frac{3\lm^2 g_A \bn}{16m\F} \frac{1}{E-m} \frac{1}{2}
(1+\gz)+\cdots,\\
\Pi (E)_{(h)} &=&- \frac{3\lm^2 g_A^2 \bn}{16\F} \frac{1}{(E-m)^2} \frac{1}{2}
(1+\gz) +\cdots
\eea
where the ellipses denote non-pole terms. Again observe that the graph (h)
does not give rise to a simple pole.

Collecting the results for the simple and the double poles at $E=m$, we find
the vacuum pole (3.5) to be modified in nuclear medium to
\[ -\frac{\lm^{* 2}}{E - m^* + i\ep}\,\frac{1}{2} (1+\gz),\]
where
\bea
\lm^* &=& \lm\left \{ 1-\left ( \frac{3g_A}{8m\F} -2\zeta \right)\bn
\right \},\\
m^* &=& m+\frac{3}{4} \left ( C_S -C_T +\frac{g_A^2}{4\F} \right )\bn
\eea
The constant $(C_S-C_T)$ can be related directly to experiment, but no such 
relation appears to exist for $\zeta$. As is known already \cite{Montano} and we 
shall also see below, the formula for $m^*$ is unacceptable at normal nuclear 
density. 

\section{Modified Formula}
\setcounter{equation}{0}
\renewcommand{\theequation}{4.\arabic{equation}}

If we note that the nucleon mass-shift formula (3.14) is given by graphs (b) and 
(h) of Fig.1 and recall the presence of the mass-shell delta-function in the
density dependent part of the nucleon propagator (3.3), it is easy to guess
that the coefficient of $\bar{n}$ in this formula must be related to
some on-shell $N\!N$ scattering amplitude at threshold. To obtain the actual
relation, we evaluate the scattering amplitude in the same chiral
Lagrangian framework as that Sect. II. In the tree approximation the
contributing Feynman graphs are shown in Fig.2.


\begin{figure}
\begin{center}
\includegraphics[width=12cm,height=4cm]{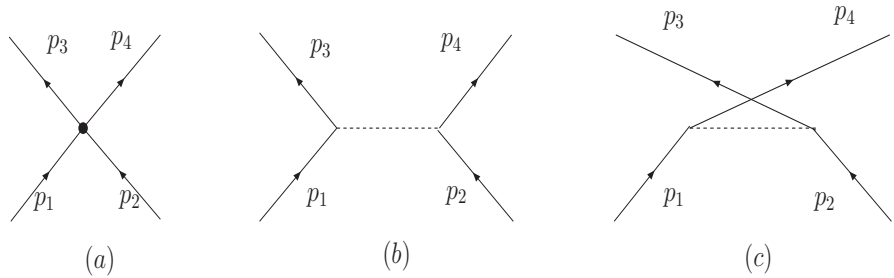}
\end{center}
\caption{Nucleon-nucleon scattering amplitude in the tree approximation.}
\end{figure}

We choose to calculate the spin averaged amplitude in the forward direction,
\[ \bM (p_1,p_2 \rw p_1,p_2) =\frac{1}{4} \sum_{\sg_1,\sg_2} M
(p_1,\sg_1;p_2,\sg_2 \rw p_1,\sg_1;p_2,\sg_2)\,.\]
The propagating nucleon, say, a proton, may scatter with a proton or a
neutron in the medium. Thus the isospin structure of $\bM$ in a
symmetric medium must be given by 
\[ \bM = \bM_{pp\rw pp} +\bM_{pn\rw pn} \,.\]
With this spin and isospin structure, the amplitude $\bM$ as calculated
from the graphs of Fig.2 is $(Q=p_2-p_1)$,
\bea
\bM=&\frac{1}{2}&\sum_{i=S,T} C_i\{ -tr\Gamma_i(\ps_1+m)\Gamma_i(\ps_2+m)
+2tr\Gamma_i(\ps_1+m)\cdot tr \Gamma_i(\ps_2+m)\} \nonumber \\
& + & \frac{3}{4} \left (
\frac{g_A}{2\F} \right )^2 tr(\ps_1+m)\Qs\gf (\ps_2+m)\Qs\gf /(Q^2-m_\pi^2) \nonumber \\
 = &-& 6m^2\left \{ (C_S-C_T)\frac{(E_1+m)(E_2+m)}{4m^2} +
\frac{g_A^2}{4\F} \right \}\,,
\eea
$E_1$ and $E_2$ being the energies of the two nucleons. (The graph (b) in Fig.2 
does not contribute to $\bM $.) Thus at threshold $\bM $ involves the same 
combination of constants as those in Eq.(3.14) for $m^*$. 

As shown in Appendix B, the amplitude $\bM $ at threshold can be expressed in 
terms of the s-wave spin-singlet and -triplet scattering lengths $a_1$ and $a_3$ 
to get
\be
m^* -m =\frac{3\pi}{2m}(a_1 +a_3) \bn\,.
\ee
With their experimental values as quoted in the Appendix and at normal nuclear
density ($p_F=1.36\, fm^{-1}$), it gives $m^* -m=-620 $ MeV \cite{Montano}, 
which is unacceptably large.

To look for the source of the problem, it is suggestive to rewrite (3.14)
and (4.2) as
\be
m^* - m = -\frac{1}{m}\int^{p_F}_0 \frac{d^3p}{(2\pi)^3 2m}
\left \{-6m^2\left ( C_S -C_T +\frac{g_A^2}{4\F}\right)\right\}
=-\frac{1}{m}\int^{p_F}_0 \frac{d^3p}{(2\pi)^32m}
\left\{-12\pi m (a_1+a_3)\right\}
\ee
At this point we recall Weinberg's method for the two-nucleon amplitude. The 
constants in the curly brackets in the integrands above represent the scattering 
amplitude in terms of the s-wave scattering lengths. Clearly this approximation 
is too bad in the low energy region, as it does not reproduce the nearby poles 
due to the bound and virtual states. But we can treat these constants as effective
potentials (to lowest order) and solve the Lippmann-Schwinger equation with
it to get a better representation of the partial wave amplitudes. As Weinberg 
himself shows \cite{Weinberg2}, the solution in this case turns out to be the 
unitarized version of the potential. Accordingly we replace the constants by 
the corresponding momentum dependent amplitudes satisfying (elastic) unitarity,
\be
-a_i \rw f^{(1)}_i(k) =\left( -\frac{1}{a_i} -ik\right)^{-1},~~~i=1,3\,.
\ee 
Here $k$ is the centre-of-mass momentum, to be distinguished from $p$, which
is the momentum of the in-medium nucleons in the rest frame of the nucleon
under consideration, the two being related by $k^2=m(\sqrt{m^2+p^2}-m)/2$.

An even better approximation of the scattering amplitude is the effective range 
approximation, that would result from including derivative terms in the
effective theory \cite{Kaplan}. The corresponding replacement would read
\be
-a_i \rw f^{(2)}_i(k) =\left( -\frac{1}{a_i} +\frac{1}{2} r_i k^2
-ik\right)^{-1},~~~i=1,3\,,
\ee 
where $r_{1,3}$ are the effective ranges in the spin-singlet and -triplet 
s waves. Without trying to relate the effective ranges to the coupling
constants of the higher order terms in the effective Lagrangian, we shall 
take their values as well as those of the scattering lengths from experiment.

We thus get both the real and the imaginary parts of the pole position as, 
\be 
m^* - \frac{i}{2}\gm = m -\frac{6\pi}{m}\int^{p_F}_0 \frac{d^3p}{(2\pi)^3} 
\{f_1(k) + f_3(k)\}\,.
\ee
where $f_i(k)$ stands for $f^{(1)}_i(k)$ or $f^{(2)}_i(k)$, according as we
choose the replacement (4.4) or (4.5). The width $\gm$ is to be interpreted 
as the damping rate of excitations with quantum numbers of the nucleon.
The numerical evaluation of Eq.(4.6) is shown in Figs. 3 and 4. In normal
nuclear matter, we get $\dm \equiv m^* -m =-33$ MeV and $\gm/2 = 110$ MeV, 
with amplitudes in the effective range approximation (solid curves in Figs.3 
and 4).

\vspace{0.5cm} 
%

\begin{figure}
\begin{center}
\includegraphics[width=8cm,height=5cm]{epjmass.eps}
\end{center}
\caption{Nucleon mass shift in nuclear matter as a function of Fermi
momentum. The dashed and the solid curves are drawn with amplitudes
$f_i^{(1)}(k)$ and $f_i^{(2)}(k)$ respectively.}
\caption{Nucleon mass shift in nuclear matter as a function of Fermi
momentum.}
\end{figure}

\begin{figure}
\begin{center}
\includegraphics[width=8cm,height=5cm]{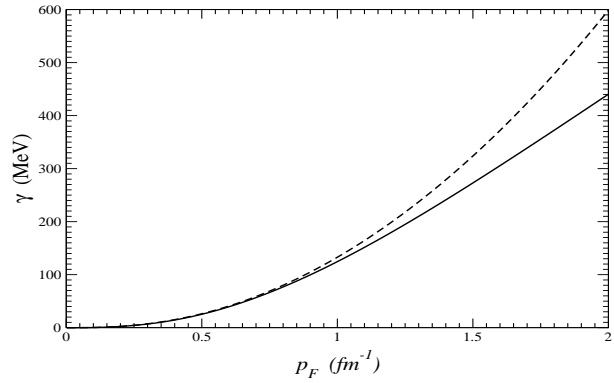}
\end{center}
\caption{Nucleon width in nuclear matter as a function of 
Fermi momentum. The dashed and the solid curves are drawn with the same 
amplitudes as in Fig.3.}
\end{figure}

Our results may be compared with those in the literature. There have been a
number of calculations of the nucleon spectral function in terms of the
(off-shell) $N\!N$ scattering amplitude evaluated with available $N\!N$
potentials. We pick out from these works the real and the imaginary parts of
the on-shell nucleon self-energy in normal nuclear matter at zero three
momentum. Thus Baldo et al \cite{Baldo} give $\dm=-55$ MeV, $\gm/2 = 63$
MeV (as reproduced in \cite{Lehr}). Benhar et al give $\dm=-68$ MeV,
$\gm/2 = 16$ MeV. Jong et al give $\dm=-65$ MeV, $\gm/2 = 25$ MeV. Also a
phenomenological determination of the width in terms of the differential
cross-section for the $N\!N$ scattering gives $\gm/2 > 25$ MeV \cite{Lehr}.
Finally we compare with a calculation by us \cite{Mallik1} based on the
virial formula for the nucleon pole position, which gives $\dm=-37$ MeV,
$\gm/2 = 112$ MeV.

\section{discussion}
\setcounter{equation}{0}
\renewcommand{\theequation}{5.\arabic{equation}}

The spectral function of nucleon in nuclear matter depends on the interaction
in a two-nucleon system. To treat this interaction in the framework of
effective chiral field theory \cite{Weinberg2}, one has to derive first the
effective potential from the effective Lagrangian and then solve the
dynamical equation with this potential. In this way one can restore the
otherwise invalid power counting rule and accomodate at the same time the
singularities of the scattering amplitude in the low energy region. 

The failure of the earlier calculation of the effective nucleon mass
\cite{Montano} may now be understood in this framework as due to representing
the two-nucleon scattering amplitude by the (leading term of the) effective 
potential itself. In this work we proceed further to carry out the next
dynamical step, replacing the potential by the unitarised scattering amplitude. 
By including the effective range terms in the s-wave amplitudes, we actually 
include the next-to-leading order terms also in the effective potential.

It will be noted that our mass-shift and width formulae do not involve
any Pauli blocking effect, our improved scattering amplitude being still in 
{\it vacuum} and not in {\it medium}. An estimate of this effect can be made 
from the work of Ref. \cite{Machleidt}, where one finds a change of about 
$20\%$ in the value of the potential energy per nucleon in nuclear matter, 
when the Pauli projection operators are withheld from the calculation. 
Incidentally, the phenomenological formula for the width in Ref.\cite{Lehr} 
involves the differential cross-section in {\it vacuum}, the Pauli blocking 
factors appearing only for the final nucleons. Such factors, however, do not 
appear in our formula, as it depends on the forward scattering amplitude, where 
the final particles occupy the states vacated by the initial ones.

The formula (4.6) bears a close resemblance to the one obtained from the virial
expansion of the nucleon self-energy to first order \cite{Mallik2}. In the virial 
formula, it is the full (spin averaged) forward amplitude that appears in the 
integral, which was evaluated in Ref.\cite{Mallik1} using the phase shift
analysis of experimental data on $N\!N$ scattering \cite{Nijmegen}, while
the integral here includes only the part of this amplitude corresponding to
$s$-waves in the effective range approximation. But in the low energy region
over which the integrals are evaluated, this approximation for the $s$-waves
agrees well with the phase shift analysis. Further, the higher partial waves
contribute negligibly to the virial formula. We thus expect the close
agreement of the present results with those from the virial formula,
stated at the end of the previous Section. 
 
\section*{appendix A}
\setcounter{equation}{0}
\renewcommand{\theequation}{A.\arabic{equation}}

Here we describe a method to evaluate the density dependent part of the
amplitude for the one-loop graphs. In this method we calculate 
the imaginary part of the Feynman amplitude and construct its real part by a
dispersion integral over it. Since we exclude the vacuum contribution and
consider only the density dependent part, no subtraction is necessary for
these integrals. The same result can be obtained in a simpler way by
retaining only the density dependent part of the nucleon propagator in the 
amplitude and using the mass-shell delta function in it to integrate out the 
energy component of the four-momentum. But the dispersion method is more
transparent in that it shows the cut structures, where the contributions come from.

We now  evaluate graph (e) of Fig.1 in some detail, it being the prototype
of all other graphs. It has the amplitude
\be
\Pi(q)_{(e)}=-\frac{3\lm^2}{4\F}\Gm(q)
\ee
where
\be
\Gm(q)_{11}=i\int d^4 x e^{iq.x} \frac{1}{i} D(x)_{11}\gf \frac{1}{i}
S(x)_{11}\gf.
\ee
Due to the presence of $\theta (\pm p_0)$ in Eq.(3.3) for $S(p)_{11}$, it is
not convenient to transform the amplitude in momentum space. Instead we
integrate out the $p_0$ variable in $S(x)_{11}$ itself, getting
\bea
\frac{1}{i}S(x)_{11}&=&\int\frac{d^3p}{(2\pi)^3
2\omp}\Big[\left\{(\ps+m)(1-n^-)e^{-ip.x}
+(\ps-m)n^+e^{ip.x}\right\}\theta(x_0) \nonumber \\
&-&
\left\{(\ps+m)n^-e^{-ip.x}+(\ps-m)(1-n^+)e^{ip.x}\right\}\theta(-x_0)\Big],~~~
p_0\equiv \omp \,.
\eea
The analogous expression for the pion propagator is
\be
\frac{1}{i} D(x)_{11}=\int\frac{d^3k}{(2\pi)^32\omk}\Big[\left\{(1+n)e^{-ik.x}+
n e^{ik.x}\right\}\theta(x_0) +\left\{ne^{-ik.x}+(1+n) e^{ik.x}\right\}
\theta(-x_0)\Big],~~~k_0\equiv \om_k \,
\ee
where $n$ is the pion distribution function, 
$ n = (e^{\bet \omk} -1)^{-1},~~ \omk=\sqrt{m_\pi^2+k^2}$.
Though $n^+(\omp)$ and $n(\omk)$ are zero for the medium we are interested 
in, we retain them at this stage for generality and symmetry.

With these expressions for the propagators, we may carry out both the $x^0$ 
and $\vec x$ integrations in Eq.(A.2) giving the energy denominator and the
3-momentum delta function respectively. The imaginary part may now be read
off and put in the form,
\be
Im\Gm(q)_{11}=\pi \tanh (\bet (q_0 -\mu)/2)Im \Gm (q),
\ee
where
\be
Im \Gm(q)=
-\int\frac{d^3p}{(2\pi)^32\omp}\int\frac{d^3k}{(2\pi)^32\omk}
(\ps-m)\left\{(1-n^-+n)\delta^{(4)} (q-p-k)+(n^-+n)\delta^{(4)} (q-p+k)\right\}.
\ee
Here the first term corresponds to $\eta\rw\pi N$ and the second 
to $\eta + \pi \rw N$, giving rise respectively to the discontinuity across 
the unitary and the `short' cuts \cite{Weldon}. With $q_0=E$ and for $\vq=0$, 
at which we work, these cuts extend in the $E$-plane over $E\geq m+m_\pi$ and 
$0\leq E \leq m-m_\pi$. Note the opposite sign before $n^-$ in the two terms.
 
The complete result for $Im \Gm$ has also another piece given by 
\[-\int\frac{d^3p}{(2\pi)^32\omp}\int\frac{d^3k}{(2\pi)^32\omk}
(\ps+m)\left\{(1-n^++n)\delta^{(4)} (q+p+k)+(n^++n)\delta^{(4)} (q+p-k)\right\},\]
which is non-zero for $E<0$ only and has no term proportional to $n^-$.
Clearly this observation does not depend on the structure of vertices in 
the loop diagrams. We thus have the general result that there is no 
spectral function to one loop for $E<0$ in a medium with nucleons only.

The 3- momentum integrations in Eq.(A.6) can be carried out immediately 
to get
\be
Im\Gm (E) = \pm \frac{f(E)}{E}, ~~~~~~
 f(E) = \frac{\sqrt{\om^2-m^2}}{8\pi^2} (\gz \om -m) n^- (\om),
\ee 
where $\om= (E^2 +m^2 -m^2_{\pi})/2E~$ is the nucleon energy $\omp$, as
restricted by the delta functions. The $\pm$ signs correspond to the unitary 
and the short cuts respectively. Inserting this result in Eq.(3.2), we get the 
spectral representation for $\Gm (E)$,
\be
\Gm (E)= \int_{m+m_\pi}^\infty \frac{dE'f(E')}{E'(E'-E)}
-\int_0^{m-m_\pi}\frac{dE'f(E')}{E'(E'-E)}\, .
\ee
The range of the second integral can be mapped on to that of the first by 
the inverse transformation $E' \rw (m^2-m_\pi^2)/E'$. Noting that $\om$ 
and hence $f(E)$ is form invariant under this transformation, we have
\be
\Gm (E)=\int_{m+m_\pi}^\infty\frac{dE'}{E'}f(E')
\left(\frac{1}{E'-E}-\frac{1}{(m^2-m_\pi^2)/E' -E}\right) .
\ee
Once we are on the unitary cut, the kinematics is determined by the first
$\de$-function in Eq.(A.6). For $\vq=\vec 0$, it gives 
\be
E'=\sqrt{m^2+p^2}+\sqrt{m_\pi^2+p^2},
\ee
so that $E'$ and $p$ are the total energy and the 3-momentum in the  
centre-of-mass frame of the intermediate $\pi N$ system. The distribution function 
restricts the upper
limit of the integral to the energy given by Eq.(A.10) with $p$ replaced by the
Fermi momentum $p_F$. Now setting $m_\pi =0$, Eq.(A.9) gives $\Gm$ to leading 
order as
\be
\Gm (E)=-\frac{1}{4\pi^2}\int_0^{p_F} \frac{dp\, p^2}{(E-m)^2 -p^2} (1-\gz)\,,
\ee
which is finite at $E=m$.

\section*{appendix B}
\setcounter{equation}{0}
\renewcommand{\theequation}{B.\arabic{equation}}

The result (4.1) corresponds to s-wave amplitudes. To get the exact
combination of these amplitudes, we recall the partial wave analysis of 
$\bM $. Following the usual convention, we write it in the center-of-mass 
system as
\be
\bM=8\pi W\left( \frac{3}{2}\obf^{(I=1)}(k)
+\frac{1}{2}\obf^{(I=0)}(k)\right)\,,
\ee
where $W$ and $k$ are the total energy and momentum in the c.m. frame. Each 
of the isospin amplitudes above has the partial wave expansion \cite{Weinberg3},
\be
\obf^{(I)}(k) = 2.\frac{1}{4}\sum_{j,s,l} (2j+1) f^{Ijs}_{l,l}(k)\,.
\ee
Here the factor $2$ is due to the identity of the scattering particles.
The total angular momentum $j$ is obtained by coupling the total spin
and orbital angular momenta $s$ and $l$. In general, $f^{Ijs}$ is a  
$2\times 2$ matrix in $l$ space, whose diagonal elements enter the sum in
Eq.(4.3). The quantum numbers of the partial wave amplitudes are governed by
the antisymmetry of the total wave function,
\[(-1)^l\cdot (-1)^{1-s}\cdot (-1)^{1-I} = -1\,.\]
Thus in the s-wave approximation, Eq.(4.2) reduces to 
\be
\bM=6\pi W\{f_1(k) +f_3(k)\} \stackrel{threshold}{\longrightarrow} 
-12\pi m (a_1 +a_3),
\ee
where $f_{1,3}$ are the spin singlet and triplet s-wave amplitudes and $a_{1,3},$ the
corresponding scattering lengths. Experimentally $a_1=-23.74$ fm and
$a_3=5.31$ fm \cite{Bohr}. We also need the experimental values for the
effective ranges, $r_1 = 2.7$ fm and $r_3 =1.70$ fm \cite{Bohr}. 

\section*{Acknowledgments}

One of us (S.M.) is grateful to Professor H. Leutwyler for helpful 
suggestions. He also acknowledges earlier support of CSIR, Government of
India. The other (H. M.) wishes to thank Saha Institute for Nuclear Physics, 
India for warm hospitality.

\end{document}